\documentclass[journal, twocolumn]{IEEEtran}

\IEEEoverridecommandlockouts

\usepackage{cite}
\usepackage{amsmath,amssymb,amsfonts}
\usepackage{gensymb}
\usepackage{float}
\usepackage{stfloats}
\usepackage{array}

\usepackage{algorithmic}
\usepackage{graphicx}
\usepackage{textcomp}
\usepackage{xcolor}
\usepackage[hidelinks]{hyperref}
\usepackage{cancel}
\usepackage{pgf}

\usepackage{amsthm}
\theoremstyle{definition}

\ifCLASSOPTIONcompsoc
    \usepackage[caption=false, font=normalsize, labelfont=sf, textfont=sf]{subfig}
\else
\usepackage[caption=false, font=footnotesize]{subfig}
\fi

\begin{document}

\title{Sizing Antenna Arrays for Near-field Communication and Sensing}

\author{
    Marcin Wachowiak,~\IEEEmembership{Member,~IEEE,} 
    André Bourdoux,~\IEEEmembership{Senior~Member,~IEEE,}
    Sofie Pollin,~\IEEEmembership{Member,~IEEE,}%
    
    \thanks{
        Marcin Wachowiak and Sofie Pollin are with imec, 3001 Leuven, Belgium and also with the Katholieke Universiteit Leuven, 3000 Leuven, Belgium (e-mail: marcin.wachowiak@imec.be)}%
    \thanks{
        André Bourdoux is with imec, 3001 Leuven, Belgium. (Corresponding author: \textit{Marcin Wachowiak})}
}

\maketitle

\begin{abstract}
This paper presents key performance metrics for near-field communication and sensing systems and their scaling behavior as a function of the antenna array aperture. 
Analytical expressions are derived for several standard array geometries to ease the design of the large antenna arrays under given 
system requirements. First, the near-field beam focusing is analyzed and the minimum beamdepth is observed to rapidly saturate to a low asymptotic limit as the array aperture increases. 
In contrast, the near-field region span is shown to scale quadratically with the array aperture. Based on these two metrics, the maximum number of resolvable beamspots at 3 dB separation is derived analytically, 
exhibiting a linear dependence on the array aperture. 
Moreover, when considering a region where the beamfocusing resolution does not exceed a specified threshold, the extent of the region is also shown to scale linearly with the array size. 
Finally, the number of significant singular values of a channel observed at the array's broadside
is estimated, showing a power-law dependence on the aperture. 
The resulting expressions provide practical design guidelines for evaluating aperture requirements in near-field communication and sensing applications. 

\end{abstract}


\section{Introduction}

\subsection{Problem Statement}

Extremely large antenna arrays are seen as the next step in the evolution of wireless communication systems, following the massive MIMO. 
Large radiating aperture facilitates an extended radiative near-field (NF) region in which the propagation is governed by the spherical wave model. In the radiative near-field, each antenna of the array observes a slightly different phase shift to some point of interest, which is dependent on both range and angle, which are coupled.
The main property of the radiative near-field systems is that the array factor (AF) is a function of both angle and distance from the array, unlike in the far-field, where it is solely a function of angle. 
This facilitates beamfocusing, also known as range-dependent beamforming. 
In the communications context, the additional phase diversity enables multiplexing of the users in the range domain \cite{nf_comms}. From the sensing perspective, the near-field allows separation of targets in range without bandwidth, only relying on the angle diversity offered by the spherical wavefront \cite{nf_sensing}.

The extent of the near-field region and its benefits are primarily a function of the array aperture and the wavelength. However, there is little agreement on the size and center operating frequency of the future near-field systems. This highlights the need for a parametric analysis of the near-field systems to translate the near-field sensing and communication performance metrics into clear antenna array design criteria.

\begin{figure}[t]
    \centering
    \includegraphics[width=\linewidth]{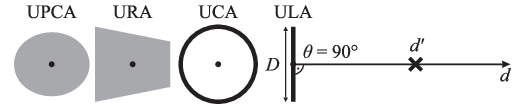}
    \caption{System diagram.}
    \label{fig:sys_diagram}
\end{figure}

\subsection{Relevant works}
The extent of a near-field region and the beamdepth of several array geometries have been discussed in \cite{primer_on_nf_bf, large_arr_beamdepth, uca_near_field}. Listed papers investigated the communication performance of the apertures for some arbitrarily selected array sizes and operating frequencies. 
The near-field sensing performance is primarily evaluated in terms of Cramér-Rao lower bounds (CRLB) \cite{crlb_nf_sensing}, which are suitable for a single target but cumbersome to use. A cooptimization of the sensing and communication performance in terms of CRLB can be found in \cite{nf_jcns}.

\subsection{Contributions}
In this work, the near-field communication and sensing performance is quantified by several new metrics such as the minimum beamdepth, span of the region with guaranteed resolution, maximum number of $3$ dB separated beamspots and number of significant singular values in the near-field region. The metrics are expressed as a function of the array aperture, which is then expressed as a multiple of the wavelength, simplifying the analysis to a single variable. The analysis is performed for several popular array geometries, namely uniform linear (ULA), circular (UCA), rectangular (URA) and planar circular arrays (UPCA). The provided closed-form expressions allow one to easily determine the minimum aperture size required to satisfy specific sensing and communication design criteria.

\section{System model}
\begin{figure}[b]
    \centering
    \includegraphics[width=\linewidth]{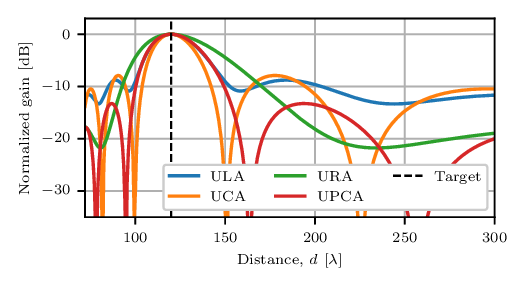}
    \caption{Array factor per geometry for $D=60\lambda$ and beamfocusing at $d'= 120 \lambda$.}
    \label{fig:af_per_geom}
\end{figure}
Consider a narrowband antenna array with aperture $D$, defined as the maximum extent of the geometry. The array is operates at a center frequency $f_{\mathrm{c}}$, corresponding to wavelength $\lambda = \mathrm{c}/f_{\mathrm{c}}$, where $\mathrm{c}$ is the speed of light. The element spacing satisfies the Nyquist criterion and is half-wavelength, i.e, $\lambda/2$. The antenna elements are assumed to be isotropic and mutual coupling is considered negligible.

In the radiative near-field region, the array factor (AF) of an antenna array is a function of the distance $d$ and can be written as $\mathrm{AF}\left( x(d', d) \right)$, where $d'$ is the beamfocusing distance, which corresponds to the target or user location. In the near-field, the spherical distance difference across antennas is approximated using a quadratic second-order Taylor approximation. As a result, even for different geometries, the argument of the AF shares the same formulation  \cite{primer_on_nf_bf, large_arr_beamdepth, uca_near_field} given by
\begin{align}
    \label{eq:nf_af_arg}
    x(d', d) &= a d_{\mathrm{FA}} d_{\mathrm{ver}}
\end{align}
where $a$ is an argument scaling coefficient dependent on the array geometry \cite{mw_nf_res_alpha} and $d_{\mathrm{ver}} = \left| \frac{1}{d'}- \frac{1}{d} \right| = \frac{|d - d'|}{d'd}$ is the absolute value of vergence difference \cite{geom_optics}.
The beamforming resolution is evaluated in terms of the $-3$ dB beamwidth, which in the range is referred to as the beamdepth (BD). The 3 dB BD is calculated by computing the argument for which the AF power drops to half of its maximum value $|\mathrm{AF}(x_{\mathrm{3 dB}})|^2 = 0.5$ and then solving for $d$ in \eqref{eq:nf_af_arg}, yielding 
\begin{align}
    \label{eq:3dB_distance}
    d_{\mathrm{3 dB}} = \frac{d_{\mathrm{FA}} d'}{ d_{\mathrm{FA}} \pm \alpha d'}.
\end{align}
Formula \eqref{eq:3dB_distance} gives two distances for which the $-3$ dB value is achieved when beamfocusing at $d'$ \cite{primer_on_nf_bf}. Where $\alpha = x_{\mathrm{3 dB}} / a$ represents the normalized argument $d_{\mathrm{FA}} d_{\mathrm{ver}}$ for which the AF reaches half of the maximum power.
Given the universal formulation, the parameter $\alpha$ provides a unified measure of the beamfocusing resolution across array geometries.
Table \ref{tab:alpha_per_geom} lists the numerical $\alpha$ values obtained in \cite{mw_nf_res_alpha}. Note that the $\alpha_{\mathrm{MIMO}}$ parameter is obtained from combining the array factors of the transmit and receive apertures corresponding to the monostatic MIMO sensing scenario. It is not applicable to MIMO in the context of communications. In the following analysis and numerical results, $\alpha_{\mathrm{SIMO/MISO}}$ is used.

In this work, the array beamfocusing performance in range is assessed based on the same aperture size, which results in a different number of antennas per geometry.
Tab. \ref{tab:nant_per_geom} compares the number of antennas per geometry required to achieve the same aperture.
In the following, consider the target or user positioned on the array normal (broadside) within the radiative near-field of the array. Fig. \ref{fig:sys_diagram} illustrates the system setup. The broadside case corresponds to the best-case scenario, due to the largest effective aperture, resulting in the best achievable performance metrics. 
For angles other than the broadside, the effective aperture size is reduced and can be approximated as a projection of the array on a plane perpendicular to the axis of the target \cite{large_arr_beamdepth}. For ULA, the effective aperture for a target at azimuth angle $\theta$ is $D_{\mathrm{eff}} = D \sin{(\theta)}$. For UCA, the aperture is constant with regard to $\theta$ due to circular symmetry of the array. For planar arrays, the effective aperture along each dimension is dependent on the target elevation angle and can be approximated as $D_{x, \mathrm{eff}} = D_x \sqrt{ 1 - \cos^2{(\theta)}\sin^2{(\phi)}}$, $D_{y, \mathrm{eff}} = D_y \sqrt{ 1 - \sin^2{(\theta)}\sin^2{(\phi)}}$,
where $D_x$ and $D_y$ are the array aperture extent in the X and Y axes and $\phi$ is the target elevation angle.

\begin{table}[t]
    \caption{Parameter $\alpha$ per array geometry and configuration.}
    \label{tab:alpha_per_geom}
    \centering
    \def\arraystretch{1.2}
    \begin{tabular}{
      >{\centering\arraybackslash}m{0.20\linewidth}<{}
      |>{\centering\arraybackslash}m{0.18\linewidth}<{}
      |>{\centering\arraybackslash}m{0.13\linewidth}<{}
}
         Array geometry & $\alpha_{\mathrm{SIMO/MISO}}$ & $\alpha_{\mathrm{MIMO}}$ \\
         \hline
         ULA & 6.952 & 4.969 \\
         \hline
         UCA & 5.737 & 4.148 \\
         \hline
         URA & 9.937 & 7.068 \\
         \hline
         UPCA & 7.087 & 5.103 \\
         
    \end{tabular}
\end{table}
\begin{table}[t]
    \caption{Number of antennas per geometry for $\lambda/2$ spacing.}
    \label{tab:nant_per_geom}
    \centering
    \begin{tabular}{
      >{\centering\arraybackslash}m{0.075\linewidth}<{}
      |>{\centering\arraybackslash}m{0.15\linewidth}<{}
      |>{\centering\arraybackslash}m{0.1\linewidth}<{}
      |>{\centering\arraybackslash}m{0.225\linewidth}<{}
      |>{\centering\arraybackslash}m{0.2\linewidth}<{}
    }
          & ULA & UCA & URA & UPCA \\
         \hline 
         $N_{\mathrm{ant}} \rule{0pt}{3.5ex} $ & $\left\lfloor \frac{2D}{\lambda} \right\rfloor + 1 \rule{0pt}{3.5ex} $ 
         & $\left\lceil \frac{2 \pi D}{\lambda} \right\rceil \rule{0pt}{3.5ex}$ 
         & $\left( \left\lfloor \frac{\sqrt{2}D}{\lambda} \right\rfloor + 1 \right)^2 \rule{0pt}{3.5ex}$ 
         & $\approx  \pi \left( \frac{D}{\lambda} \right)^2 \rule{0pt}{3.5ex}$
    \end{tabular}
\end{table}

\section{Nearfield performance metrics}

\subsection{3 dB beamdepth (radial resolution)}
\label{sec:3db_bd}
The beamfocusing of different practical array geometries has been well investigated and approximations of the array factor in the near-field are available \cite{primer_on_nf_bf, large_arr_beamdepth, uca_near_field}. 
Fig. \ref{fig:af_per_geom} illustrates the AF per geometry for the same aperture size.
Given the closed-form AF, the radial half-power beamwidth in the range domain can be calculated, serving as a resolution metric. Note that the considered system is narrowband, meaning that the sensing resolution offered by the bandwidth is much smaller than the one provided by near-field beamfocusing.
The 3 dB beamdepth (BD) at the array broadside is given by
\begin{align}
    \label{eq:3dB_bd_vs_d}
    \mathrm{BD}(D, d') &= \frac{2\alpha d_{\mathrm{FA}} d'^2}{d_{\mathrm{FA}}^2 - \alpha^2 d'^2}
    =\frac{4 \alpha D^2  d'^2 \lambda}{4D^4 - \alpha^2 \lambda^2 d'^2},
\end{align}
where $d'$ is the distance from the geometric center of the array. 
\begin{figure}[b]
    \centering
    \includegraphics[width=\linewidth]{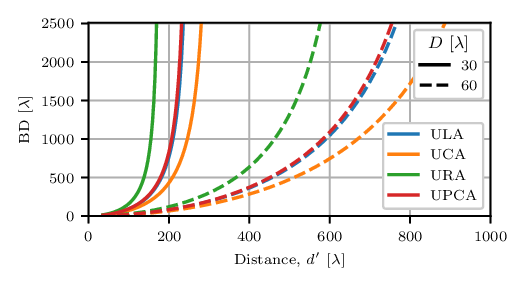}
    \caption{Beamdepth vs distance from the center of the array for two aperture sizes of $30\ \lambda$ and $60\ \lambda$, and selected array geometries.}
    \label{fig:bd_vs_d}
\end{figure}
As illustrated in Fig. \ref{fig:bd_vs_d}, the 3 dB BD is a rapidly increasing monotonic function that reaches a minimum when the distance to the array is the smallest. The aperture size determines the NF region span and the rate of increase of the 3 dB BD. Different array geometries exhibit different rates of increase due to different $\alpha$ parameters.

\begin{figure}[b]
    \centering
    \includegraphics[width=\linewidth]{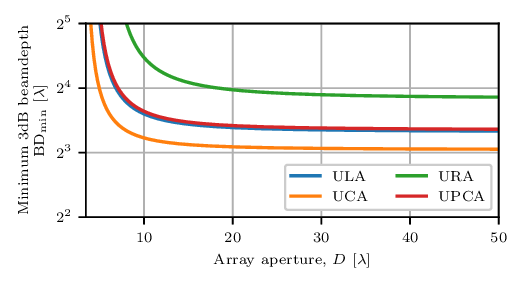}
    \caption{Minimum beamdepth $\mathrm{BD_{min}}$ vs the aperture $D$ for $\beta=1.2$ and selected array geometries.}
    \label{fig:bd_min_vs_d_ap}
\end{figure}

To calculate the minimum beamdepth consider some minimum distance from the aperture, specified as a multiple of the aperture size $d'_{\min} = \beta D$,
where $\beta$ is a scaling coefficient. Typically in the literature $\beta \geq 1.2$ \cite{primer_on_nf_bf}.
For a given minimum distance from the array, the minimum 3 dB beamdepth is
\begin{align}
    \label{eq:bd_min_vs_d_ap}
    \mathrm{BD}_{\min} = \mathrm{}\mathrm{BD}_{}(D, d'_{\min}) 
    &= \frac{4 \alpha \beta^2 \lambda}{4 - \alpha^2 \beta^2 \frac{\lambda^2}{D^2}}.
\end{align}
Fig. \ref{fig:bd_min_vs_d_ap} illustrates that for a fixed minimum distance $d'_{\min}$, the minimum achievable beamdepth quickly reaches saturation as the aperture size increases. 
Since $d'_{\min}$ scales with the array size, increasing the aperture size beyond a certain threshold does not further improve the beamfocusing resolution.
\begin{table}[t]
    \caption{Asymptotic minimum 3 dB beamdepth per geometry and configuration for $\beta=1.2$}
    \label{tab:bd_min_per_geom}
    \centering
    \def\arraystretch{1.2}
    \begin{tabular}{
      >{\centering\arraybackslash}m{0.20\linewidth}<{}
      |>{\centering\arraybackslash}m{0.27\linewidth}<{}
      |>{\centering\arraybackslash}m{0.20\linewidth}<{}
    }
         Array geometry & $\mathrm{BD_{min}^{SIMO/MISO}}$ $[\lambda]$ & $\mathrm{BD_{min}^{MIMO}}$ $[\lambda]$ \\
         \hline
         ULA & 10.01 & 7.16\\
         \hline
         UCA & 8.26 & 5.97\\
         \hline
         URA & 14.31 & 10.18\\
         \hline
         UPCA & 10.21 & 7.35\\
    
    \end{tabular}
\end{table}
As the aperture becomes infinitely large, the limit of the minimum 3 dB beamdepth is
\begin{align}
    \label{eq:asymptotic_3dB_bd}
    \lim_{D \to \infty} \mathrm{BD(D, d'_{\min})} = \alpha \beta^2 \lambda.
\end{align}
The asymptotic minimum achievable near-field resolution at a fixed distance $\beta = 1.2$ per geometry is listed in Table \ref{tab:bd_min_per_geom}.

Given the asymptotic limit and saturation behaviour of the 3 dB BD function, a key design consideration is determining the aperture size that guarantees a specified fraction of the asymptotic resolution.
The minimum aperture size $D_{\min}$ that offers $\eta > 1$ fraction of the asymptotic resolution is obtained by equating \eqref{eq:bd_min_vs_d_ap} with \eqref{eq:asymptotic_3dB_bd} and solving for $D_{\min}$ resulting in
\begin{align}
    \label{eq:min_d_ap_size}
    D(\eta)_{\min} = \frac{\lambda}{2} \alpha \beta \sqrt{\frac{\eta}{\eta - 1}}.
\end{align}
For example, for $D = 20\lambda$, all array geometries offer at least $\eta=1.1$ of the asymptotic resolution and $\eta=1.01$ for $D = 60\lambda$.

\subsection{Near-field region span}
In the context of the beamfocusing, the lower limit of the near-field region is given by the distance where the Fresnel approximation remains accurate and gain variations across the array can be considered negligible, which is $d'_{\min} = \beta D, \beta \geq 1.2$ \cite{primer_on_nf_bf}. The upper limit is the maximum distance for which the beam depth \eqref{eq:3dB_bd_vs_d} remains finite, given by $d'_{\max} = d_{\mathrm{FA}} / \alpha$ \cite{primer_on_nf_bf}.
Given the lower and upper limits, the near-field region span is
\begin{align}
    \label{eq:nf_range_vs_dap}
    R_{\mathrm{NF}} 
    &= D \left( \frac{2D}{\alpha \lambda} - \beta\right).
\end{align}
The near-field region span increases quadratically with the aperture size.

\subsection{Span of the guaranteed resolution region}
\begin{figure}[b]
    \centering
    \includegraphics[width=\linewidth]{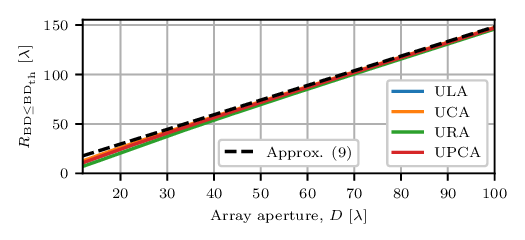}
    \caption{Span of the guaranteed resolution region as a function of the aperture $D$, for threshold resolution set to 5 times the minimum asymptotic resolution $\mathrm{BD_{th}} = 5 \alpha \beta^2 \lambda$. }
    \label{fig:res_span_vs_dap}
\end{figure}

As shown in Sec. \ref{sec:3db_bd}, the near-field beamdepth is range dependent and grows rapidly with distance. Due to the rapid increase of the BD, only a limited region of the NF is practically relevant for beamfocusing. Consider some threshold resolution of interest $\mathrm{BD}_{\mathrm{th}}$, which is greater than the minimum asymptotic resolution \eqref{eq:asymptotic_3dB_bd}. The NF region span where the BD is lower than or equal to the given threshold can be written as
\begin{align}
    \label{eq:guar_res_reg}
    R_{\mathrm{BD} \leq \mathrm{BD}_{\mathrm{th}}} 
    &= \beta  D \left( \frac{1}{\sqrt{ \left(\frac{\alpha \beta \lambda}{2D} \right)^2+ \frac{ \alpha \beta^2 \lambda}{\mathrm{BD}_{\mathrm{th}}}}} - 1 \right).
\end{align}
Given that the array size satisfies $D \gg \alpha \beta \lambda$, which is required to create multiple beamspots (See Sec. \ref{eq:min_dap} for details on the minimum array aperture in the NF), the \eqref{eq:guar_res_reg} can be approximated as
\begin{align}
    R_{\mathrm{BD} \leq \mathrm{BD}_{\mathrm{th}}} \approx 
    & \beta  D_{\lambda} \left( \sqrt{\frac{\mathrm{BD}_{\mathrm{th}}}{\alpha \beta^2 \lambda}} - 1 \right).
\end{align}
The guaranteed resolution region scales approximately linearly with the array aperture and with the square root of the specified resolution threshold. Fig. \ref{fig:res_span_vs_dap} presents the accuracy of the approximation and the desired resolution region span as a function of the aperture size.

\subsection{Number of 3 dB separated beamspots in the near-field}
\label{sec:n_3db_beams}

\begin{figure}[b]
    \centering
    \includegraphics[width=\linewidth]{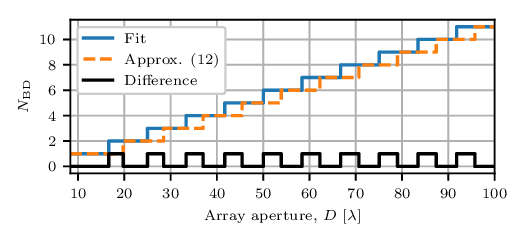}
    \caption{Maximum number of 3 dB beamspots for ULA vs the array aperture $D$ obtained by the iterative fitting method and the analytical approximation from \eqref{eq:n_3dB_vs_dap}.}
    \label{fig:n_3dB_vs_dap_apx_vs_fit}
\end{figure}

\begin{figure}[t]
    \centering
    \includegraphics[width=\linewidth]{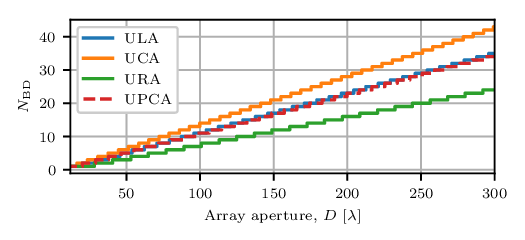}
    \caption{Maximum number of 3 dB beamspots vs the array aperture for selected array geometries.}
    \label{fig:n_3dB_vs_geom}
\end{figure}

\begin{figure}[b]
    \centering
    \includegraphics[width=\linewidth]{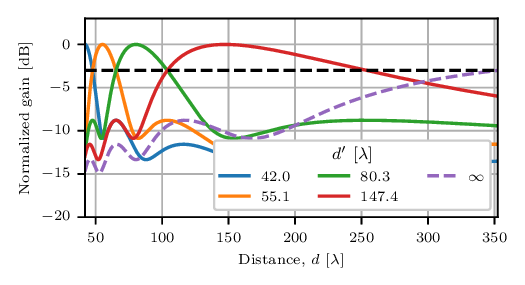}
    \caption{Array factor of the fitted 3 dB beamspots for $D = 35 \lambda$. The black dashed line specifies the $-3$ dB. The purple dashed line denotes the array factor for a far-field point. The path loss is not included in the figure.}
    \label{fig:ula_bd_af_fit}
\end{figure}

Given the 3 dB beamdepth function \eqref{fig:bd_vs_d} and an interval specifying the near-field span $d'_{\max} - d'_{\min} \geq 0 $, the number of 3 dB separated beamspots fitting within the NF span can be approximated as
\begin{align}
    N_{\mathrm{BD}} \approx\int_{d'_{\min}}^{d'_{\max}} \frac{1}{\mathrm{BD}(D, d')} \, d d'.
\end{align}
This approximation is valid for smooth and slowly varying functions, provided that the function value is small relative to the total interval. 
To accurately calculate the total number of beamspots, the beamspots at the limits of the near-field range need to be taken into account.
The first beamspot is focused at the minimum distance from the array, being $\beta D$, and occupies a near-field span corresponding only to the right half of the beamdepth. 
The near-field beamdepth is asymmetrical with regard to the $d'$ and is calculated based on the two points that can be referred to as the left side and right side, where the array factor reaches $-3$ dB of the maximum value \cite{primer_on_nf_bf}. The two points are given by 
\begin{align}
    \mathrm{BD} &= \underbrace{\frac{d_{\mathrm{FA}} d'}{d_{\mathrm{FA}} + \alpha d'}}_{d'_{\mathrm{L\ -3 dB}}} - \underbrace{\frac{d_{\mathrm{FA}} d'}{d_{\mathrm{FA}} - \alpha d'}}_{d'_{\mathrm{R\ -3 dB}}}.
\end{align}
The lower limit of the integration is the sum of the minimum distance from the array and the width of the right side of the beam allocated at the minimum distance, which results in $d'_{\min} = \frac{2 \beta D^2}{2D - \alpha \beta \lambda}$. The upper limit of the integration is limited by $d'_{\mathrm{L\ -3dB}}$ of the far-field beampattern which is $\lim_{d \to \infty} \frac{d_{\mathrm{FA}}d}{ d_{\mathrm{FA}} + \alpha d} = \frac{d_{\mathrm{FA}}}{\alpha} = d'_{\max}$ and coincides with the upper limit of the NF region.
The derived limits introduce a constraint on the minimum size of the array. To create at least one finite beam in the near-field $d'_{\max} - d'_{\min} \geq 0$, which implies that $\label{eq:min_dap} D \geq \alpha \beta \lambda$, setting a lower limit on the minimum aperture size.
Solving the integral for the given limits yields
\begin{align}
    \label{eq:n_3dB_vs_dap}
    N_{\mathrm{BD}} 
    &\approx \left\lfloor \frac{D}{\alpha \beta \lambda} + \frac{\alpha \beta \lambda}{ 4D - 2\alpha \beta \lambda} - \frac{1}{2} \right\rfloor.
\end{align}
Fig. \ref{fig:n_3dB_vs_dap_apx_vs_fit} depicts the maximum number of 3 dB beamspots obtained for ULA via the approximate solution from \eqref{eq:n_3dB_vs_dap} and iterative fitting method.
The analytical expression underestimates the maximum number of beamspots by at most one. This is due to the behaviour of the $\mathrm{BD}(D, d')$ function, which explodes for the upper limit, affecting the accuracy of the approximation. 
Fig. \ref{fig:n_3dB_vs_geom} presents the maximum number of 3 dB beamspots vs the array aperture for the considered aperture geometries. 
Fig. \ref{fig:ula_bd_af_fit} illustrates the array factor of the 3 dB beamspots created by a ULA. Note that the 3 dB separation of users or targets mitigates the main source of interference, which is the mainlobe overlap. However, the residual overlap from the sidelobes is still present and might adversely affect the performance.


\subsection{Number of significant singular values in the near-field}

\begin{figure}[t]
    \centering
    \includegraphics[width=\linewidth]{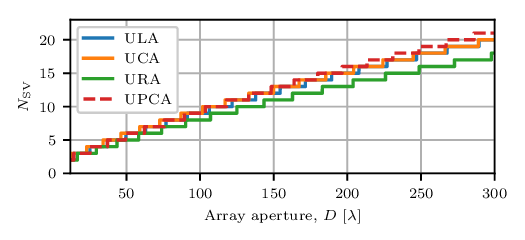}
    \caption{Number of significant singular values of the near-field channel along the array normal vs array aperture.}
    \label{fig:nsvd_vd_dap}
\end{figure}

In near-field communication and sensing, the high sidelobes of the array factor imply that the channels to the users or targets at the same angle are correlated. High correlation between channels to users at different ranges determines inter-user interference and affects target resolvability, which might be a limiting factor in near-field systems' performance. To determine how many users can be allocated at the same angle but in different ranges without interference, a singular value (SV) analysis is performed. The number of SVs is computed using extensive numerical simulations. 
Note that the recent analytical approaches do not consider the gain variations with distance and would be inaccurate in the considered scenario \cite{nsvd_los_mimo}. The channel matrix is obtained by dense sampling of the array aperture and the region of interest on the array broadside, which is then decomposed using the singular value decomposition. To determine the significant SV, a normalization with regard to the total power is applied; $\sigma_{n, \mathrm{dB}} = \sigma_n^2 / \sum \sigma_n^2 $. The number of significant SVs is defined as the number of SVs whose normalized power is equal to or greater than a threshold of $-20$ dB. This threshold captures SVs that correspond to 99\% of the total channel power.
Fig. \ref{fig:nsvd_vd_dap} presents the number of significant SV vs the array aperture. The observed dependence is not linear as one might have expected based on the $N_{\mathrm{BD}}$ analysis. The number of SVs denoted as $N_{\mathrm{SV}}$ is lower than $N_{\mathrm{BD}}$ as the channels obtained by 3 dB spacing are correlated, as can be observed by high sidelobes in Fig. \ref{fig:ula_bd_af_fit}. Moreover, the dependence on the $\alpha$ parameter is no longer explicitly visible. Based on empirical observations, the number of significant singular values can be approximated as
\begin{align}
    N_{\mathrm{SV}} = \left\lfloor \kappa \left(\frac{D}{\lambda}\right)^{\gamma}\right\rfloor,
\end{align}
where $\kappa$ and $\gamma$ are scalar parameters that depend on the array geometry. The estimated values of the $\kappa$ and $\gamma$ parameters are listed in Table \ref{tab:kappa_gamma_per_geom}.
The number of significant SVs offers insight into the number of users that can be multiplexed without interference and provides a threshold when considering the choice of precoder, depending on the number of served users.

\begin{table}[t]
    \caption{Estimated $\kappa$ and $\gamma$ per array geometry}
    \label{tab:kappa_gamma_per_geom}
    \centering
    \def\arraystretch{1.2}
    \begin{tabular}{
      >{\centering\arraybackslash}m{0.25\linewidth}<{}
      |>{\centering\arraybackslash}m{0.15\linewidth}<{}
      |>{\centering\arraybackslash}m{0.15\linewidth}<{}
    }
         Array geometry & $\kappa$ & $\gamma$ \\
         \hline
         ULA & 0.434 & 0.676\\
         \hline
         UCA & 0.503 & 0.650\\
         \hline
         URA & 0.413 & 0.662\\
         \hline
         UPCA & 0.383 & 0.707\\
    
    \end{tabular}
\end{table}

\subsection{Performance comparison}

Based on the provided results and formulas, one might be inclined to compare the performance across different geometries. However, such a comparison requires careful consideration of multiple aspects.
The provided analysis is focused on evaluating the beamfocusing performance along a single direction, at the broadside of the array. The UCA geometry offers superior and azimuth-angle-agnostic performance compared to ULA at the cost of a $2\pi$ increase in the number of antennas. Note that 1D arrays (ULA, UCA) provide beamfocusing only in range and azimuth, while the 2D arrays (URA, UPCA) are capable of beamfocusing also in the elevation, at the cost of approximately squaring the number of antennas. The URA appears to exhibit lower beamfocusing performance, owing to its aperture being defined along the diagonal. 
When comparing solely the beamfocusing performance in the boresight direction, the 2D arrays do not offer significant improvements; however, they provide improved sidelobe levels of the array factor as shown in Fig. \ref{fig:af_per_geom}. 

\section{Conclusion}

This paper presents several near-field sensing and communication performance metrics as closed-form functions of the array aperture. The unified signal model allows for a straightforward comparison across different array geometries. The minimum achievable beamdepth (resolution) is shown to quickly saturate as the aperture increases. The maximum number of 3 dB separated beamspots scales linearly and the number of significant singular values can be approximated as a power law with regard to array aperture. The provided formulas enable a straightforward translation of the system key performance indicators into array aperture requirements and vice versa.

\bibliographystyle{IEEEtran}
\bibliography{biblio.bib}

\end{document}